\documentstyle[preprint,aps]{revtex}
\begin{document}

\centerline{\bf Influence of spin-rotation measurements} 
\centerline{\bf on partial-wave analyses of elastic pion-nucleon scattering}
\vskip .5cm
\centerline{I. G. Alekseev, V. P. Kanavets, B. V. Morozov, D. N. Svirida}
\centerline{Institute for Theoretical and Experimental Physics}
\centerline{B. Cheremushkinskaya 25, 117259 Moscow, Russia}
\vskip .2cm
\centerline{S. P. Kruglov, A. A. Kulbardis, V. V. Sumachev}
\centerline{Petersburg Nuclear Physics Institute}
\centerline{Gatchina, Leningrad district, 188350 Russia}
\vskip .2cm
\centerline{R. A. Arndt, I. I. Strakovsky$^\dagger$, and R. L. Workman}
\centerline{Department of Physics, Virginia Tech, Blacksburg, VA~24061}
 
\begin{abstract}
Recent measurements of spin-rotation parameters in elastic $\pi^+ p$  
scattering are in marked disagreement with predictions of the
Carnegie-Mellon$-$Berkeley and Karlsruhe-Helsinki analyses. Using the
method of Barrelet, we show how this discrepancy can be removed.  
We then show how this Barrelet transformation alters 
the partial-wave amplitudes. The effect of unitarity and analyticity 
constraints is also considered. 
\end{abstract}

\vskip .5cm
PACS Numbers: 14.20.Gk, 13.30.Eg, 13.75.Gx, 11.80.Et

\eject

\centerline{I. Introduction}
\vskip .2cm

The excited states of the nucleon have been studied mainly through 
partial-wave analyses of $\pi N$ total, elastic and charge-exchange
scattering data. The Review of Particle Properties\cite{pdg} 
lists about 20 such states below 2 GeV. 
Many of these resonances have a 3-star or lower
rating, which implies their existence is at most ``likely'' and their
properties are uncertain. 
Most of these states were either found or verified 
in the Karlsruhe-Helsinki(KH)\cite{kh} and
Carnegie-Mellon$-$Berkeley(CMB)\cite{cmb} analyses.

In this paper we consider the effect of spin-rotation(SR) measurements
on partial-wave analyses. In particular, we focus on the KH and
CMB analyses, as no SR data were available in the resonance
region when these analyses were performed. Measurements of $R$ and $A$
between 300 MeV and 600 MeV were recently made at 
LAMPF\cite{lampf} and PNPI\cite{pnpi}. These 
quantities were at least qualitatively predicted by the existing analyses.
This was not so surprising. Earlier studies\cite{sabba} 
had suggested that the 
imposition of sufficient unitarity and analyticity constraints would 
remove those ambiguities due to the absence of SR data. However, more 
recent ITEP-PNPI measurements\cite{data} 
of $A$ for $\pi^+ p$  at 1.3 GeV {\it are} surprising.  
These data, taken between 120 and 140 degrees, suggest an angular 
dependence very different from the KH and CMB predictions. 
 
In Section II, we briefly review the problems that arise when
analyses use incomplete sets of data (no spin-rotation measurements). 
The method of Barrelet\cite{bar} 
is used to isolate and modify one zero-trajectory which
may be responsible for the poor fit to the 1.3 GeV SR data. 
The effect of a transformed trajectory on the unitarity and analyticity 
of the KH and CMB solutions is also considered. 
In Section III,
we show that the above procedure produces an improved agreement between
the SR data and the KH analysis. 
Here we also compare the partial-wave amplitudes of the original and 
modified KH solution. 
Finally, in Section IV, we summarize our results
and suggest extensions of the present work. 

\eject

\centerline{II. Amplitude Ambiguities and the Barrelet Method}
\vskip .2cm

The KH and CMB analyses were performed prior to the existence of 
spin-rotation measurements in the resonance region. These analyses 
had cross section (differential and total) and polarization
data. However, without further theoretical input, it is clear that
this dataset is insufficient. This is easier to see 
if we work with transversity
amplitudes ($F^{\pm} = F \pm i G$) constructed from the spin flip ($G$)
and spin non-flip ($F$) amplitudes. In this representation, the differential
cross section ($d\sigma / d\Omega$) and polarization ($P$) 
\begin{eqnarray}
{ {d \sigma} \over {d \Omega} } \; &=& \; |F^+|^2 + |F^-|^2 , \\
P { {d \sigma} \over {d \Omega} } \; &=& \; |F^+|^2 - |F^-|^2,
\end{eqnarray}
determine $|F^{\pm}|$, leaving an undetermined relative phase. 
(In addition to the relative phase between transversity amplitudes,
there is also an undetermined overall phase\cite{com1}.)
The total cross sections further constrain the forward scattering 
amplitudes. 

The symmetry $F^+( - \theta ) = F^-( \theta)$ can be used to express the
two transversity amplitudes in terms of a single function. Barrelet\cite{bar} 
showed that it was useful to parameterize this function as a product 
\begin{equation}
F(w) \; = \; { {F(1)}\over{w^N}} \prod_{i=1}^{2N} 
                    {{w-w_i} \over {1 - w_i}} ,
\end{equation}
in terms of the variable $w=e^{i\theta}$. The unit circle, $|w|=1$, 
corresponds to the physical region where $\theta$ is real and equal 
to the center-of-mass scattering angle. 

In writing Eq.(3), we have implicitly assumed that
$F(w)$ can be represented by a finite polynomial. 
The use of this form, in analyzing scattering data directly, has been 
criticized by H\"ohler\cite{com1}. He notes that the above product should
contain another factor, $\Re(w)$, 
which accounts for both the distant zeros
and the effects of branch cuts and poles which are known to exist but  
cannot be described by a polynomial. The violation of unitarity can also be
a problem. 

In the present work, we have applied 
the Barrelet method to the KA84\cite{kh} solution. (Results for the 
CMB\cite{cmb} and KH80\cite{kh} solutions are similar.)
Since the KH and CMB analyses have employed unitarity and analyticity 
constraints, we expect that they can be represented by the 
product given in Eq.(3), with the additional factor 
suggested in Ref.\cite{com1}. 

The operation $w_i \to 1/w_i^*$ for a single term in the above product
preserves both the cross section and polarization. This represents an  
ambiguity that can be resolved by SR measurements. When a 
zero trajectory crosses the unit circle, $w_i$ and $1/w_i^*$ are equal.
At this point, an alternate zero trajectory (with $w_i \to 1/w_i^*$) can
emerge. Whether this new trajectory connects reasonably smoothly to the 
original one also depends on the angle at which the trajectory crosses the
unit circle. 

We have identified one particular trajectory which can be linked to the   
discrepancy found in the SR measurements. This trajectory
crosses the unit circle at about 700 MeV and remains influential through
the remainder of the resonance region. When transformed,
it produces an improved description of the SR data. This trajectory  
appears in both the KH and CMB solutions. The original and transformed 
trajectories for the KA84 solution are displayed in Fig.~1. A detailed
comparison of the original and transformed solutions is given in
Section~III.

If SR data had become available at the time of the KH and CMB 
analyses, they would have been fitted along with the constraints from 
dispersion relations. We have found that the Barrelet transformation,
applied to these solutions, results in a good fit 
to SR measurements without altering the fit to the remaining database. 
If this operation has a minimal effect on the dispersion relation 
constraints, we can take the variation in partial-wave amplitudes, 
displayed in Section~III, as a guide to the results expected in 
a full partial-wave analysis. 
 
As the transformation of roots does not alter the forward amplitude, 
none of the forward dispersion-relation integrals are affected. 
Gauging the effect on other dispersion relations is more difficult, 
since we have only the $A$ measurement for elastic
$\pi^+ p$ scattering (I=3/2). A change in the I=3/2 partial-waves would
result in a readjustment of the I=1/2 amplitudes in a fit to the full
database. As a simple test, we recalculated the $\pi NN$ coupling constant 
($f^2$) using the $\pi^+ p$ amplitudes in a fixed-t dispersion relation.  
The value of $f^2$ was shifted systematically by less than 1\% over a range 
of t-values. While this change in $f^2$ is within its uncertainty, some 
small readjustment of the amplitudes, due to 
the dispersion-relation constraints, would occur in a full 
analysis\cite{com2}. 

Finally, we should note that the Barrelet transformation can result in a
violation of unitarity. This can be a serious problem if data are analyzed
directly. Here we are again helped by the fact that we {\it start} with
amplitudes which are unitary. The transformed amplitudes have been checked 
for violations of unitarity, and one example is given in the next section. 

\vskip .2cm
\centerline{III. Partial-Wave Amplitudes and Observables}
\vskip .2cm

As mentioned in Section II, the Barrelet transformation leaves invariant the
polarization and the differential and total cross sections. In Fig.~2, we
show how much better the transformed amplitudes describe the $\pi^+ p$
$A$ measurements at 1.3 GeV. This change in the prediction for $A$ starts
at about 700 MeV and persists through the resonance region. The other 
sets of $R$ and $A$ measurements just missed this effect, as they extended
up to only 600 MeV. A contour plot of differences between the predictions
of the original and transformed solutions is displayed in Fig.~3. This
transformation also removes disagreements with the analysis of 
Ref.\cite{vpi} at back angles. The KH and CMB solutions have resolved 
ambiguities in the same way\cite{borie} and, therefore, Figs.~2 and 3 
are qualitatively the same when the CMB solution is used\cite{com3}.

Differences between the original and transformed KH amplitudes are shown
in Fig.~4. Partial-waves with clear resonance signatures show the same
qualitative behavior in both solutions. Weaker resonance signals are
more significantly affected in the S, P and D waves. The higher partial
waves are again qualitatively similar in the two solutions. 

The D$_{35}$ partial wave provides an example of a case where the 
Barrelet transformation results in a small violation of unitarity.
(The imaginary part is negative.) Luckily, this problem occurs at 
energies where the imaginary part is very small. 

\vskip .2cm
\centerline{IV. Conclusions} 
\vskip .2cm

We have shown that the Barrelet transformation of one particular zero
trajectory has the effect of greatly improving the KH and CMB descriptions
of recent spin-rotation measurements. The effect of this transformation 
on the partial-wave amplitudes serves as a guide to results which would be 
expected in a full analysis. A further readjustment to satisfy unitarity   
and analyticity constraints should also be expected. This is particularly
true for the D$_{35}$ partial wave. 

This transformation, applied to the KH and CMB solutions, also 
results in an improved agreement between the KH, CMB and VPI\cite{vpi} 
predictions for $A$ in $\pi^+ p$ scattering at back angles, 
over a wide range of energies. 
It is important to have further measurements in the neighborhood
of 1.3 GeV to determine whether the modified solutions make 
improved predictions.   

There are numerous resonance candidates with masses near the center-of-mass
energy of the ITEP-PNPI measurements\cite{data}. From Fig.~4, we see 
that the well-established F$_{35}$(1905) and F$_{37}$(1950) resonances 
appear clearly in both the original and transformed solutions. Weaker
structures in the S, P and D waves have been altered significantly in
the transformed solution. This added uncertainty should be factored 
into future comparisons with quark-model predictions.

\vskip .2cm

This work was supported in part by the U.S. Department of
Energy Grant DE-FG05-88ER40454 and the Russian Fund of Fundamental
Research and Russian State Scientific-Technical Program "Fundamental
Nuclear Physics".  I.~S. acknowledges the hospitality extended by the 
Physics Department of Virginia Tech.

\newpage

%
\newpage
{\Large\bf Figure captions}\\
\newcounter{fig}
\begin{list}{Figure \arabic{fig}.}
{\usecounter{fig}\setlength{\rightmargin}{\leftmargin}}
\item
{
Comparison of the original and transformed F$^+(w)$ zero-trajectories 
which cross the unit circle near 0.7 GeV. Energy values (in GeV units)
are marked by x and z symbols on the original and transformed
trajectories respectively.  
}
\item
{
The spin-rotation parameter $A$ for $\pi^+ p$ elastic scattering 
at 1.3 GeV. The original KA84 solution (solid line)
is compared to the Barrelet-transformed solution (dot-dashed line). 
Data are taken from Ref.\cite{data}.
}
\item
{
Contour plot of differences between the modified and original KA84
solutions (see text). The white and black regions correspond to differences
less than $-0.48$ and greater than +1.15 respectively. 
Neighboring contours differ by approximately 0.2.
}
\item
{
Partial-wave amplitudes for the KA84 solution. The real (solid line) and
imaginary (long dot-dashed line) parts of the original solution are compared
to the real (dashed line) and imaginary (short dot-dashed line) parts of
the Barrelet-transformed solution. The dotted line gives the unitarity 
constraint (Im$T$-$|T|^2$) for the original solution. 
(a) S$_{31}$, (b) P$_{31}$, (c) P$_{33}$, 
(d) D$_{33}$, (e) D$_{35}$, (f) F$_{35}$, (g) F$_{37}$, (h) G$_{37}$, 
(i) G$_{39}$, (j) H$_{39}$, (k) I$_{3\;11}$.
}
\end{list}
\vfil
\eject

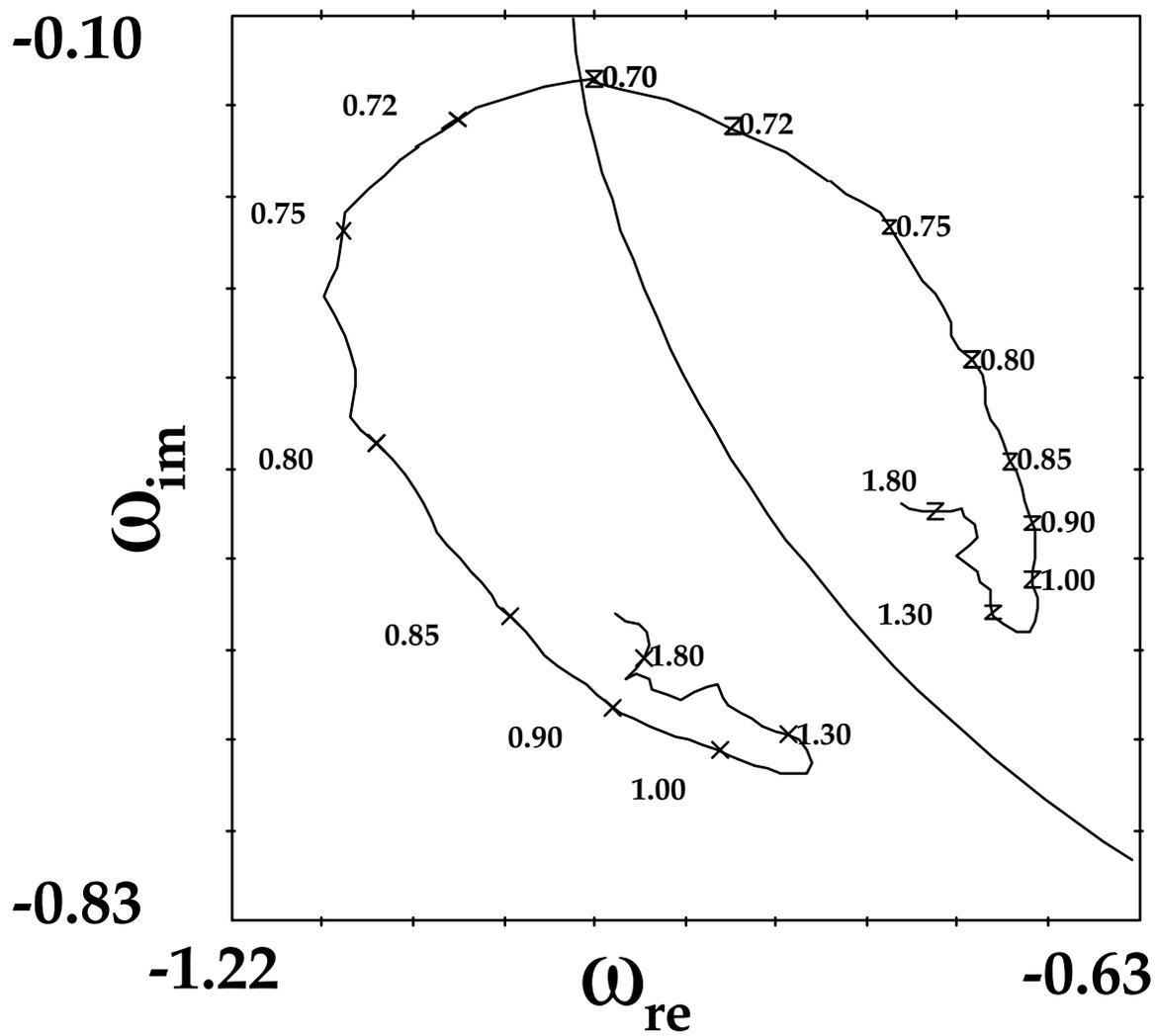
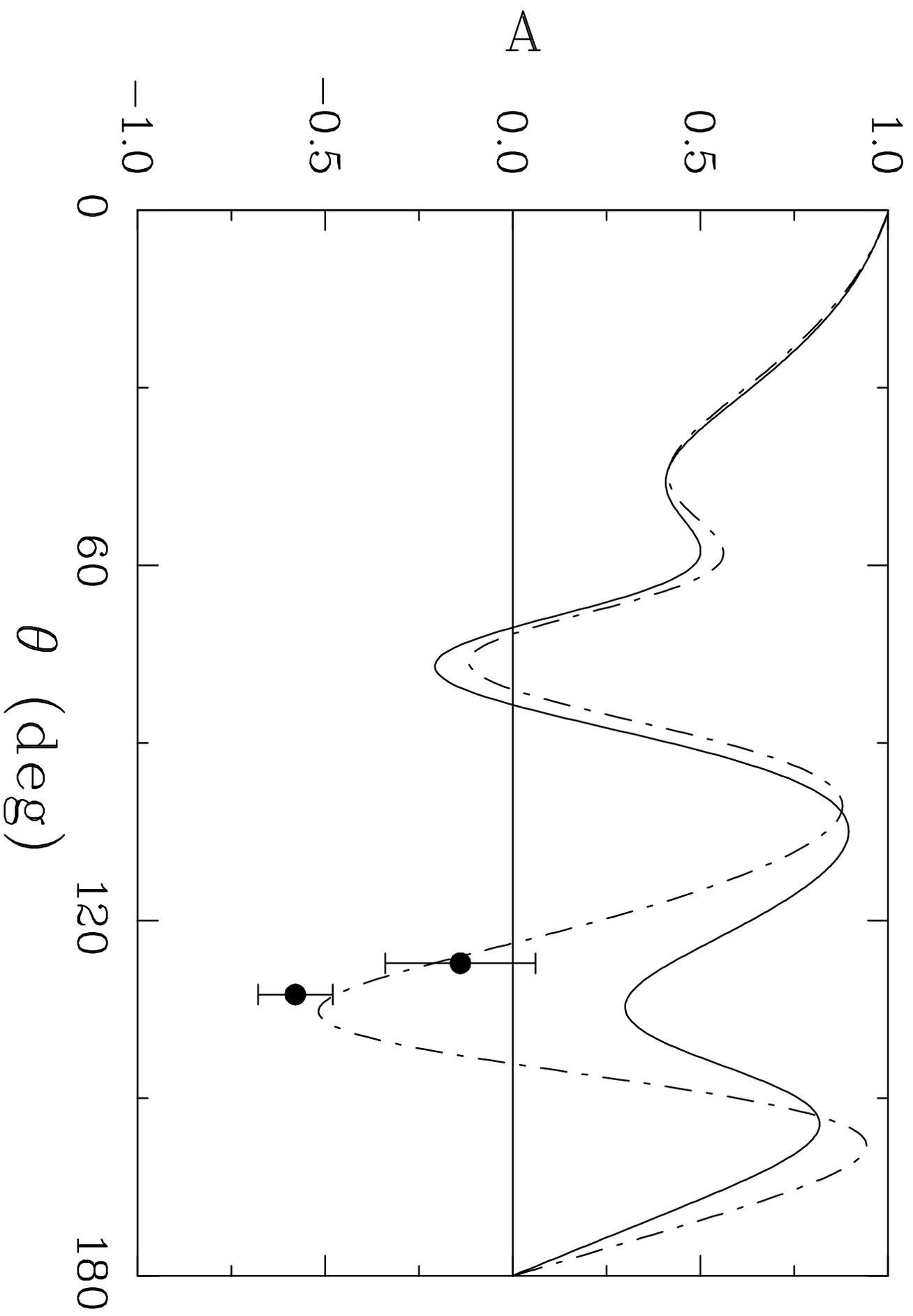
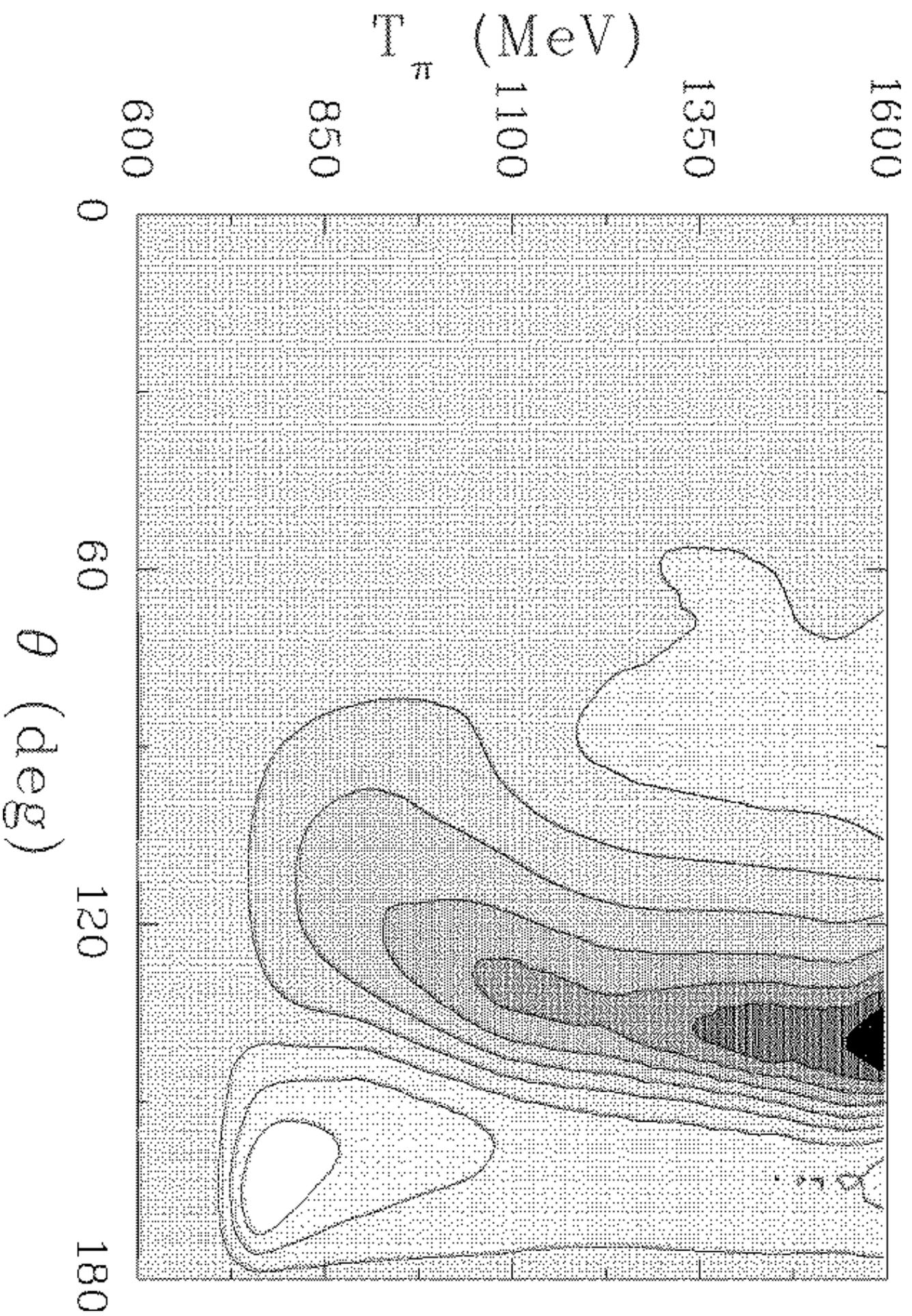
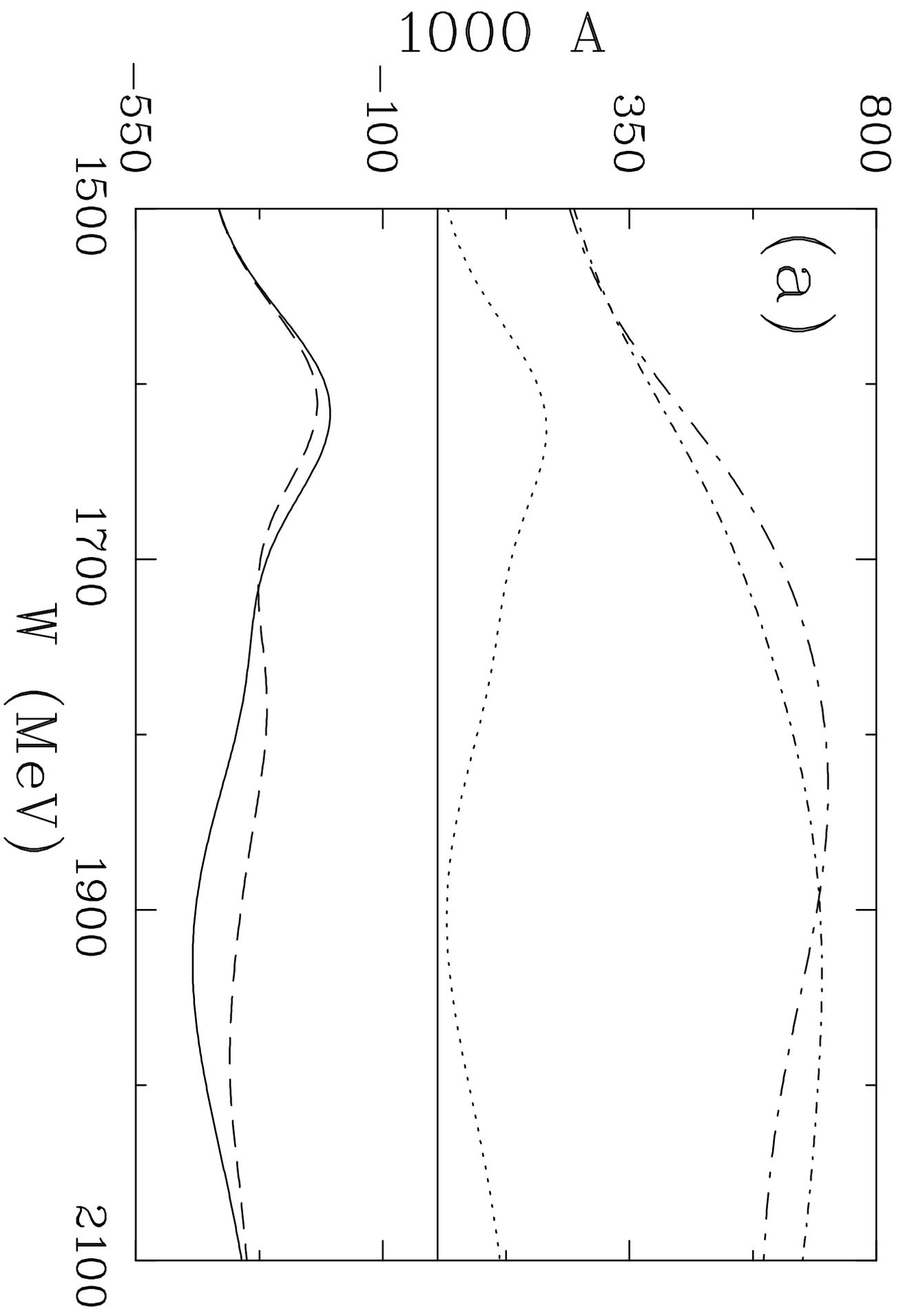
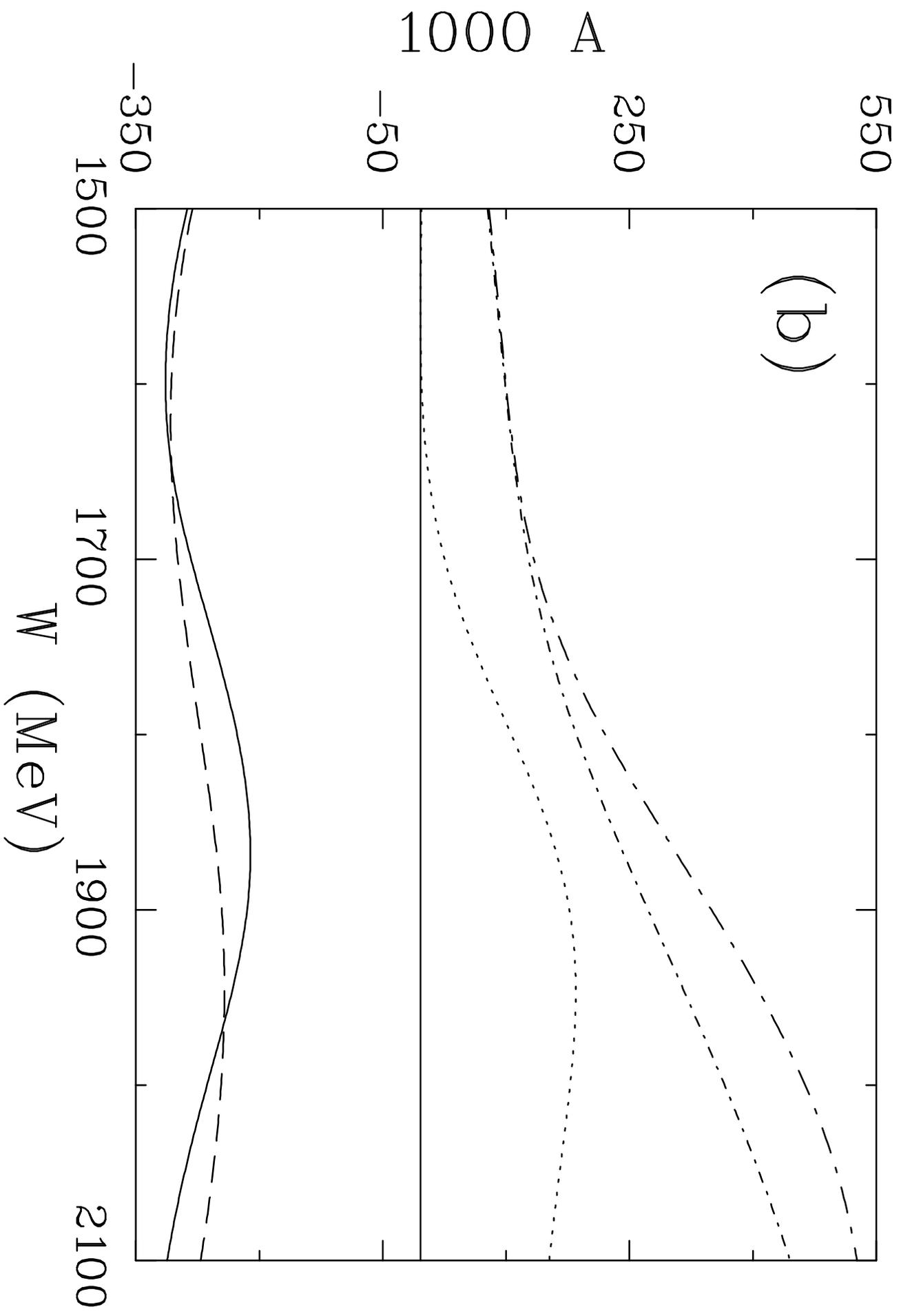
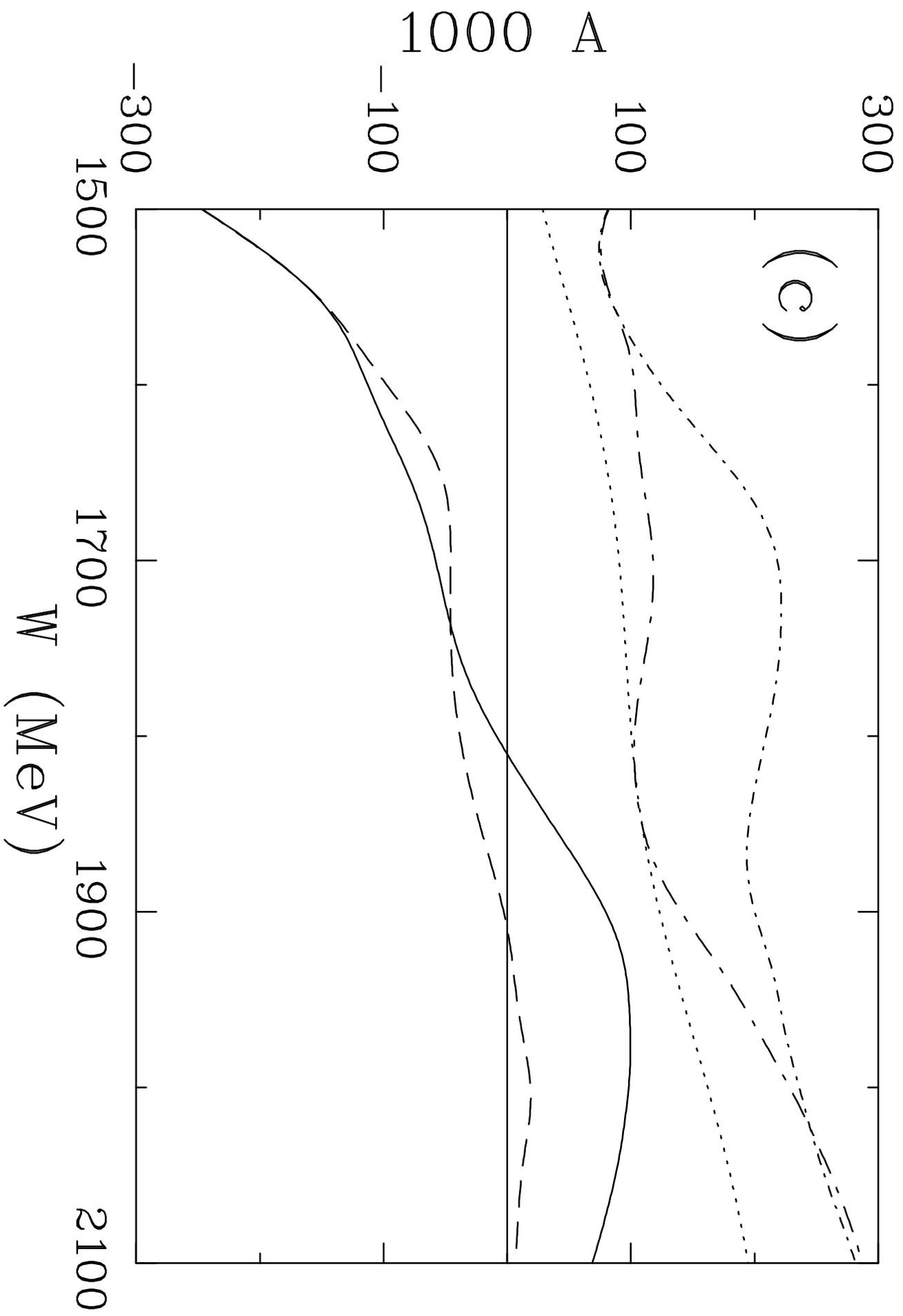
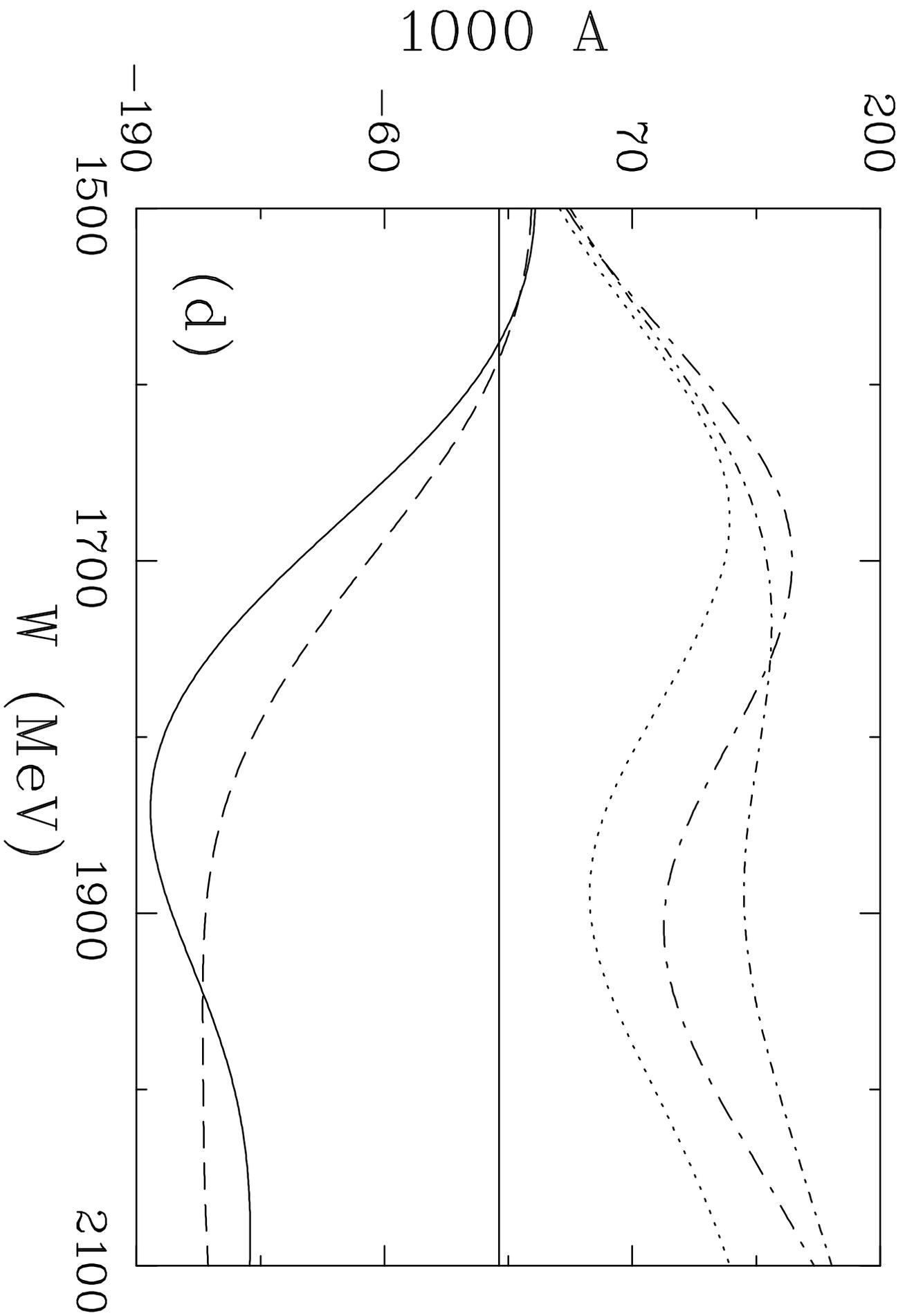
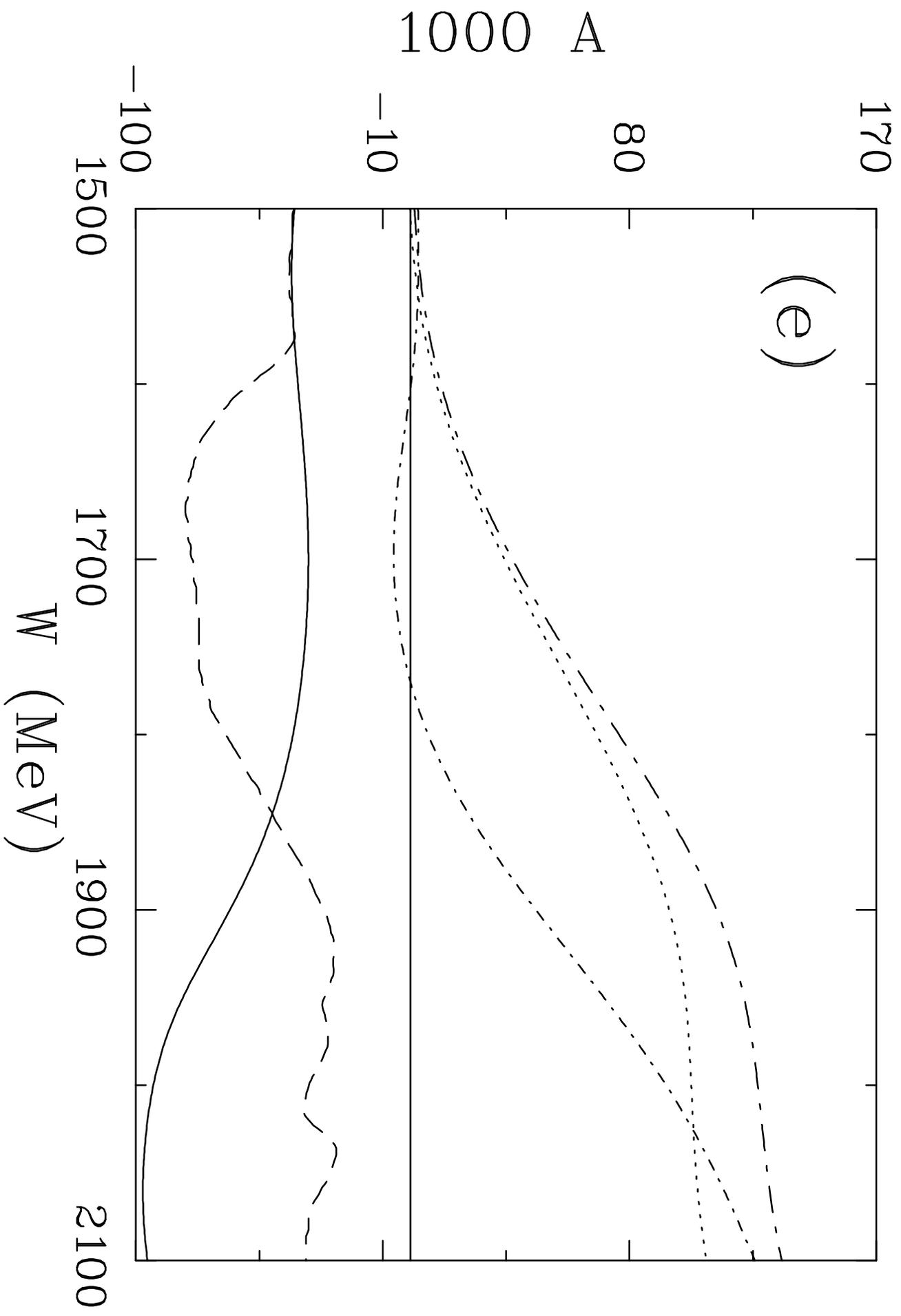
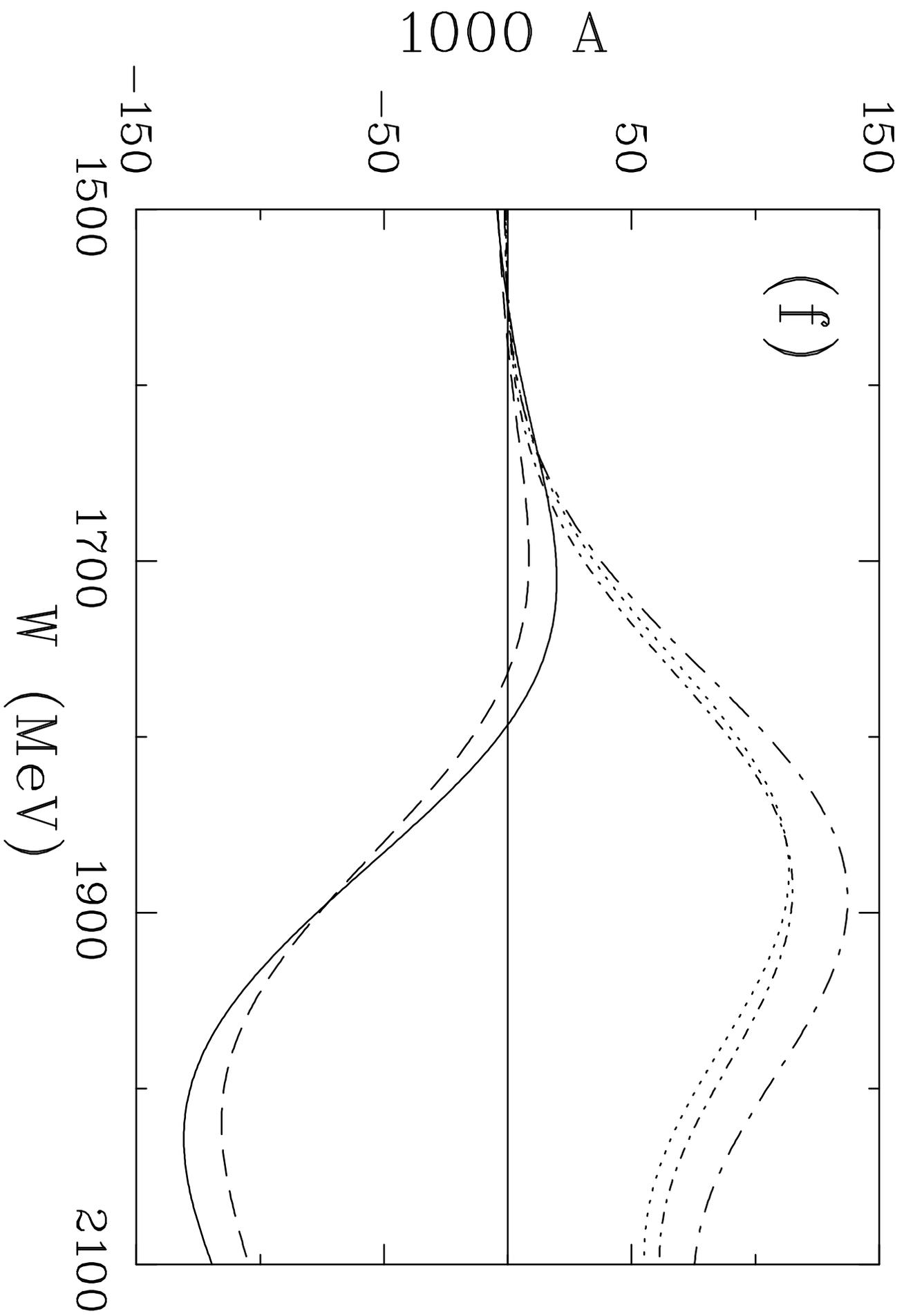
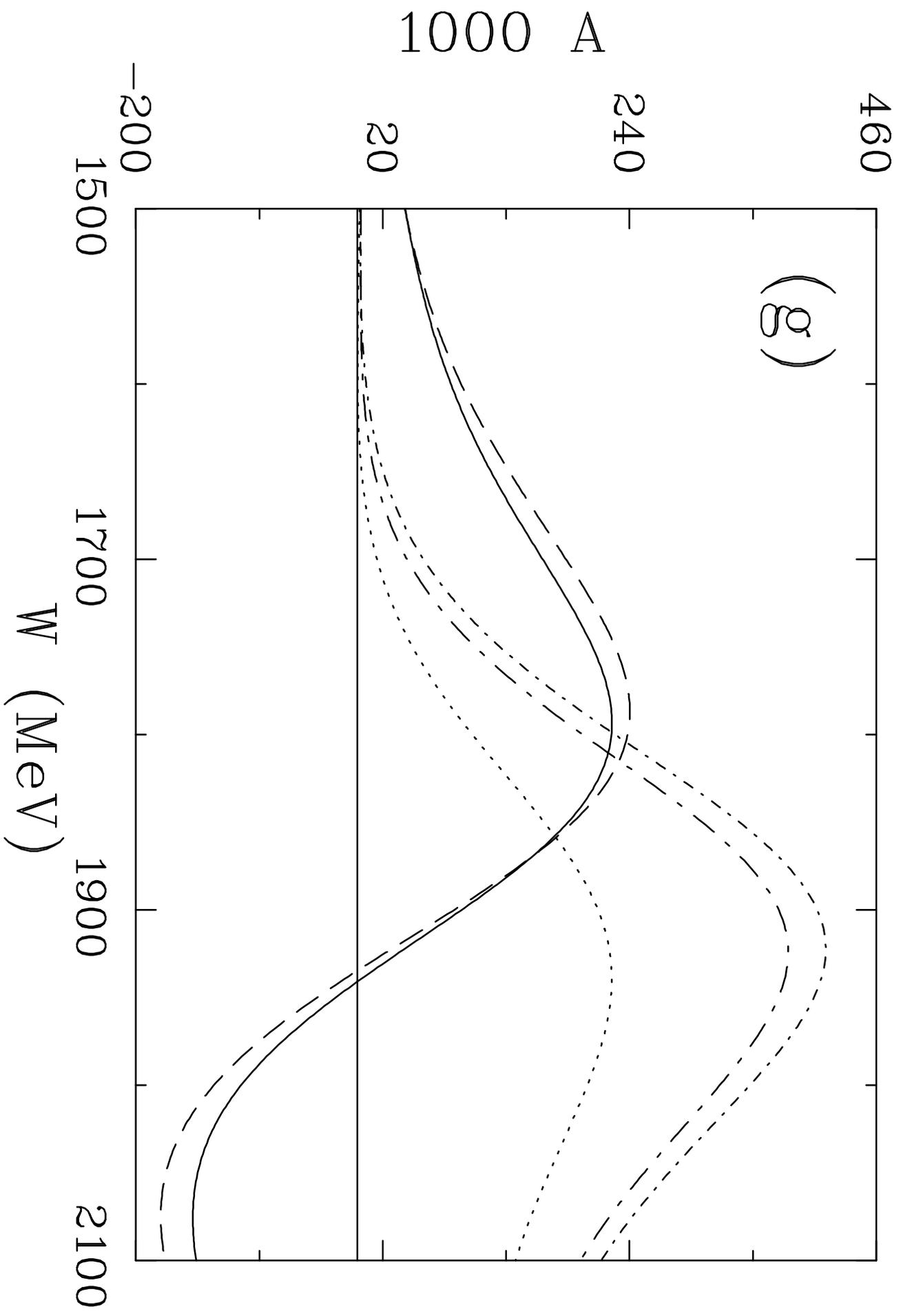
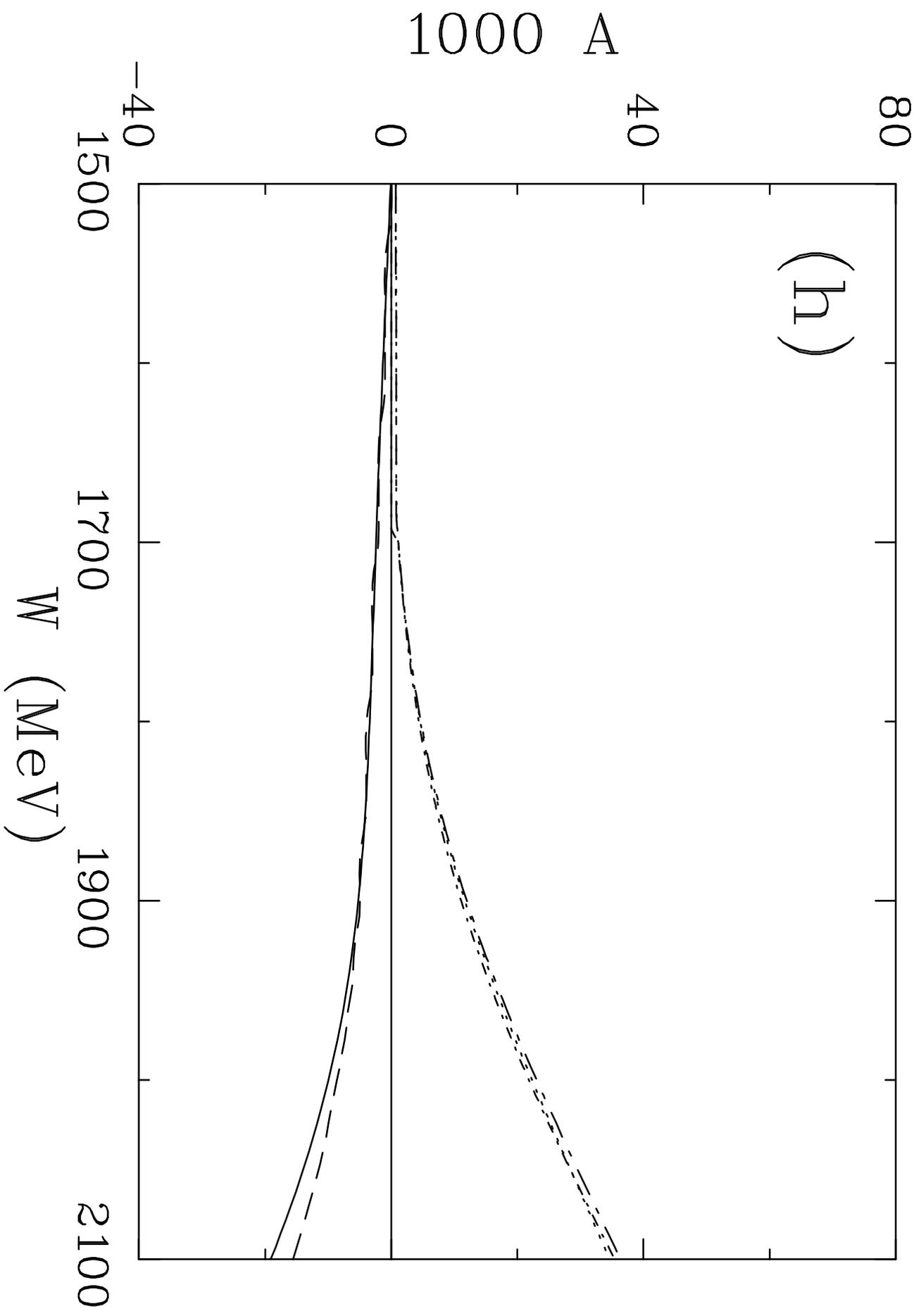
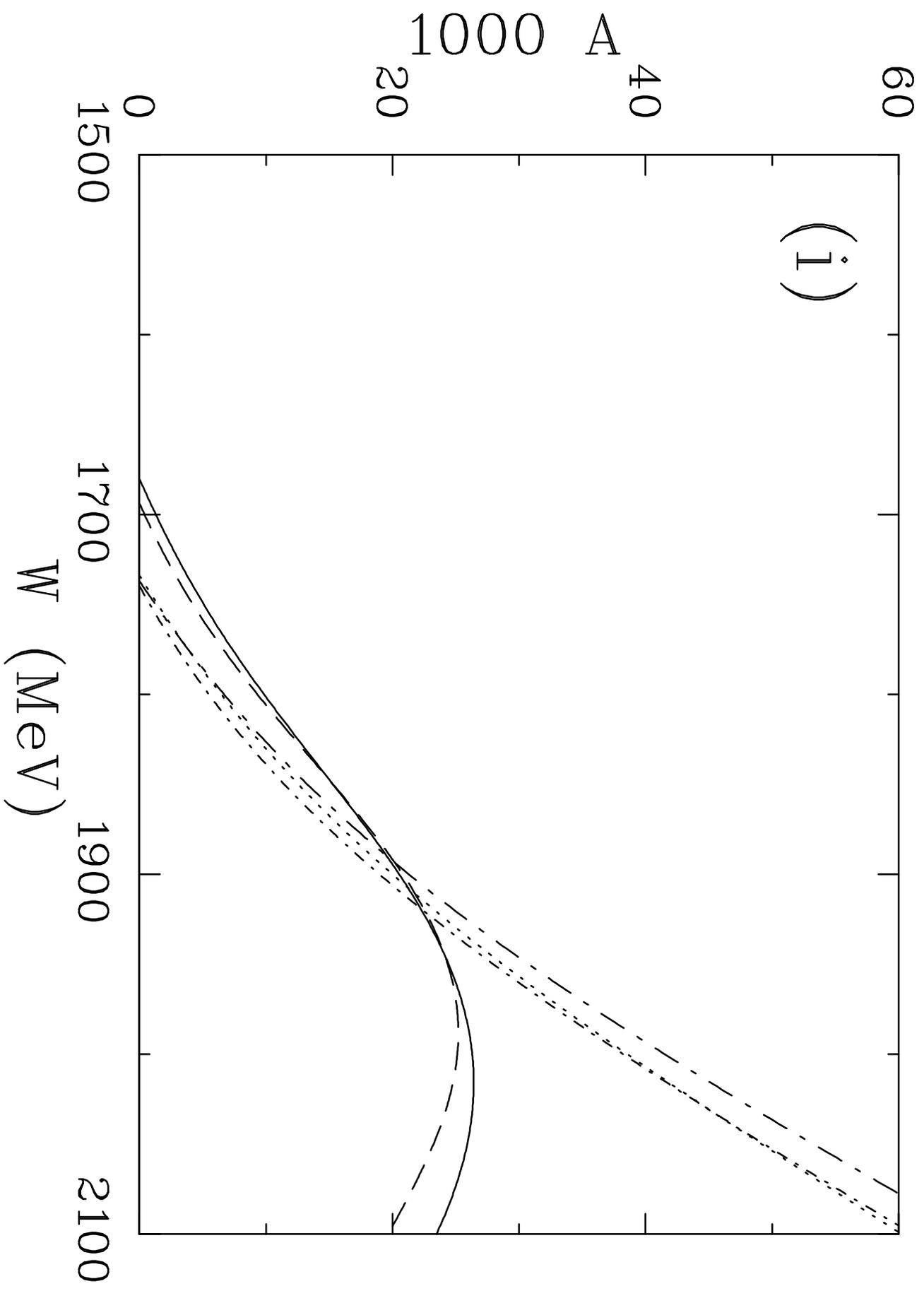
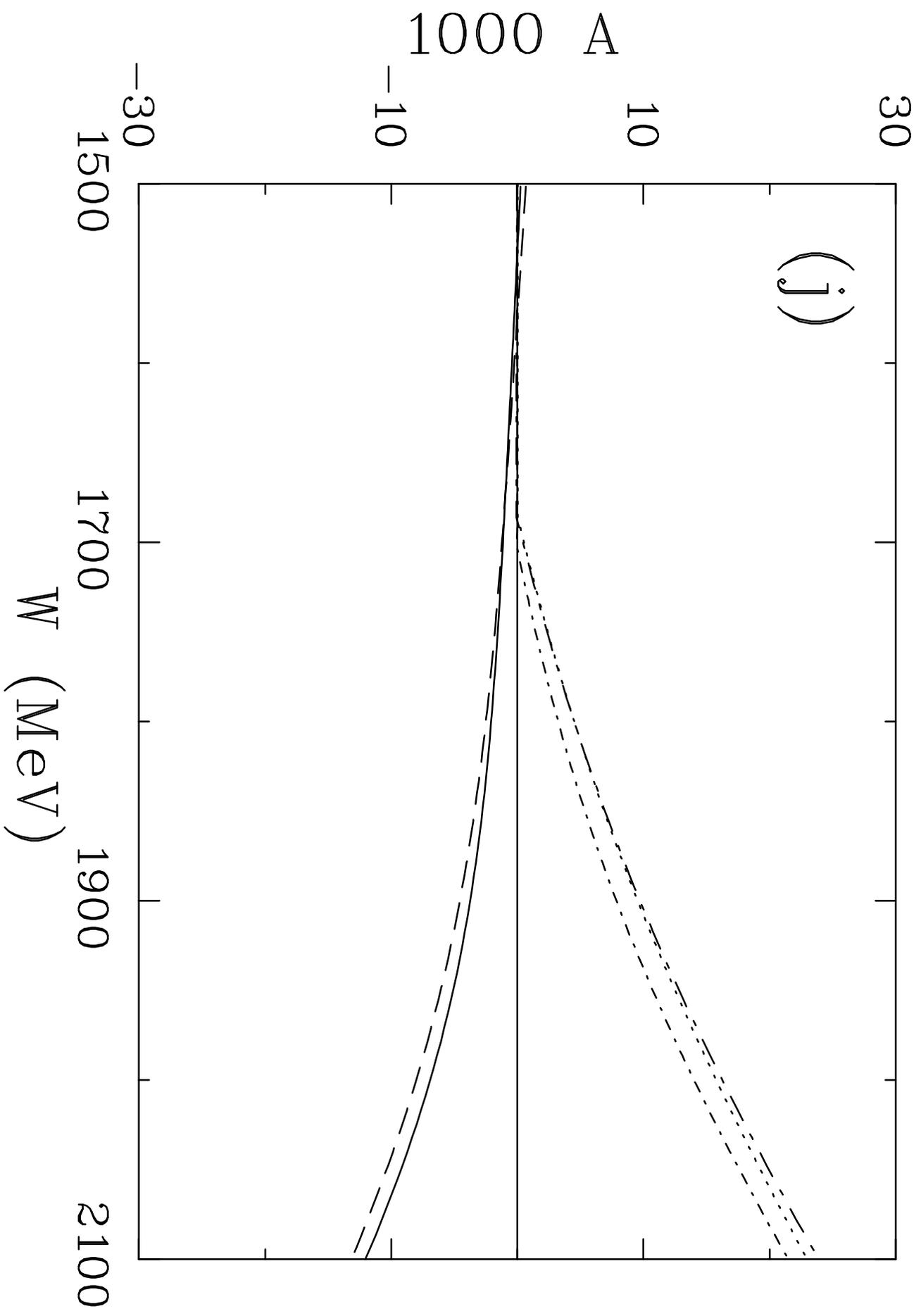
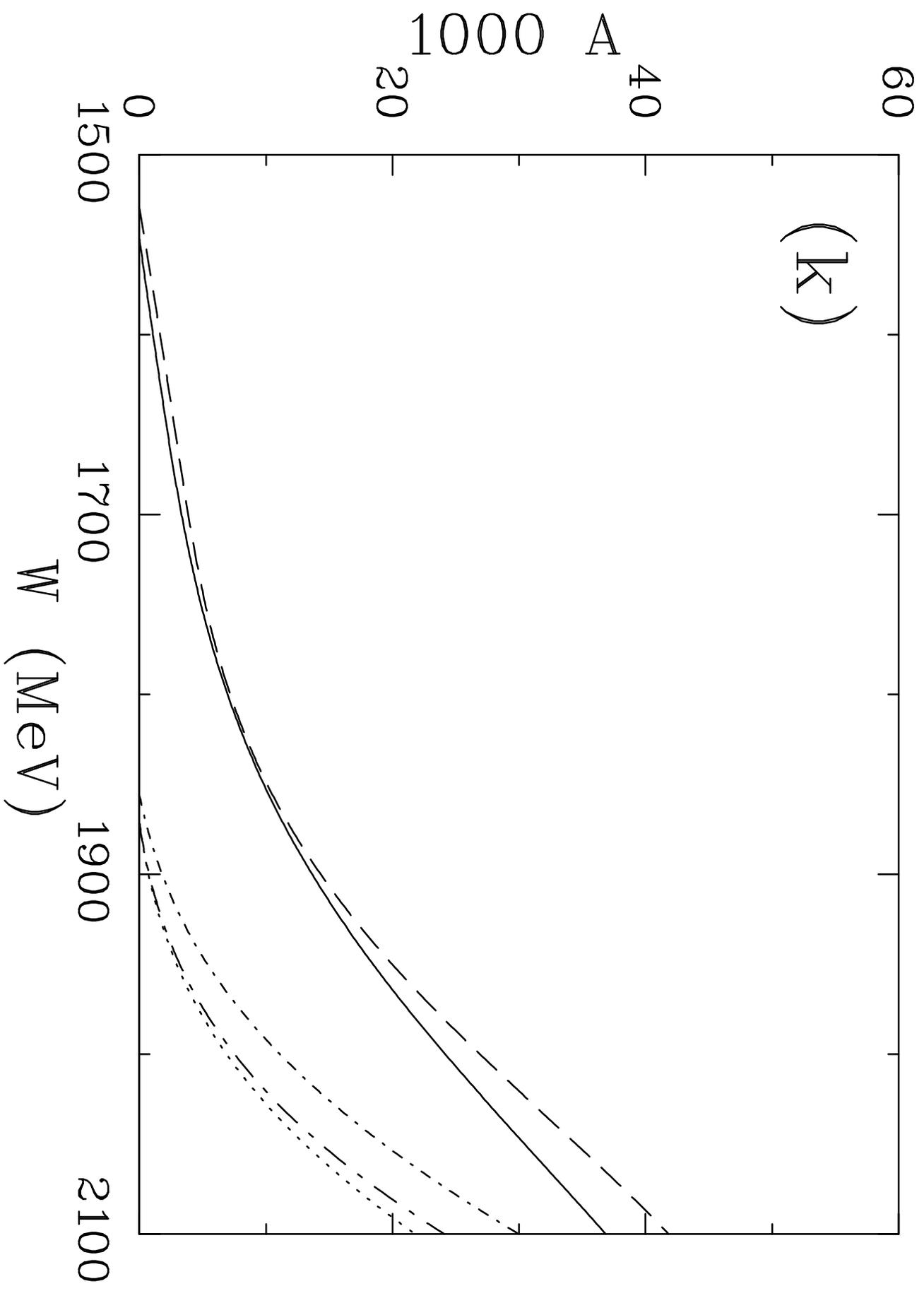
\end{document}